\documentclass[12pt]{article}
\usepackage{amsmath,amssymb,geometry,microtype}
\usepackage{marginnote,xparse,changepage,caption,graphicx,subcaption}
\usepackage[numbers,sort&compress]{natbib}
\usepackage{placeins}

\usepackage{tocloft}

\numberwithin{equation}{section}
\newcommand{\IGNORE}[1]{}
\newcommand{\be}{\begin{equation}}
\newcommand{\ee}{\end{equation}}
\newcommand{\bea}{\begin{aligned}}
\newcommand{\eea}{\end{aligned}}
\newcommand{\F}[1]{F_{#1}}
\renewcommand{\P}[1]{P_{#1}}
\newcommand{\BF}[1]{\mathbf{F}_{#1}}
\newcommand{\BP}[1]{\mathbf{P}_{#1}}

\usepackage{todonotes}
\usepackage{soul}

\RequirePackage[colorlinks=true
,urlcolor=blue
,anchorcolor=blue
,citecolor=blue
,filecolor=blue
,linkcolor=blue
,menucolor=blue
,pagecolor=blue
,linktocpage=true
,pdfproducer=medialab
]{hyperref}

\begin{document}

\vfill

\begin{center}
\baselineskip=16pt {\LARGE\bf The Geometry of Small Causal Cones}
\vskip 1.5cm {Ian Jubb}\\
\vskip .6cm
\begin{small}
\textit{
		{Theoretical Physics Group, Blackett Laboratory, Imperial College, London, SW7 2AZ, UK}\\\vspace{3pt}
		}
\end{small}
\end{center}

\vspace{20pt}

\begin{center}
\textbf{Abstract}
\end{center}

\begin{quote}
We derive a formula for the spacetime volume of a small causal cone. We use this formula within the context of causal set theory to construct causal set expressions for certain geometric quantities relating to a spacetime with a spacelike hypersurface. We also consider a scalar field on the causal set, and obtain causal set expressions relating to its normal derivatives with respect to the hypersurface.
\end{quote}

\pagebreak
\tableofcontents

\section{Introduction}

Even after 100 years of General Relativity there is still much we do not understand about spacetime and Lorentzian geometry, and of course, how this all fits together with quantum mechanics via the elusive theory of quantum gravity. One aspect of spacetime structure in which research has been fruitful recently is the geometry of certain small spacetime regions~\cite{Gibbons:2007nm,Jacobson:2015hqa,Benincasa:2010ac,Dowker:2013vba,Buck:2015oaa,Khetrapal:2012ux,Roy:2012uz,Myrheim:1978}. Understanding the geometry of such regions has led to new ways of deriving Einstein's equations from a different set of fundamental principles~\cite{Jacobson:2015hqa}, and an understanding of the geometry of small spacetime intervals, in particular, has been beneficial for one approach to quantum gravity --- causal set theory~\cite{Sorkin:2003bx,Surya:2011yh}. There is motivation, therefore, to study small spacetime regions in Lorentzian geometry, to further our understanding of spacetime and to provide tools in the search for quantum gravity.

In this paper the particular small region of interest will be the \emph{causal cone}, which will be defined shortly. In Section \ref{Volume of a Small Causal Cone} we will derive a universal formula for the volume of such a region. This formula will be general in that it can be applied to a wide class of spacetimes. In Section \ref{Use In Causal Set Theory} we will use this formula in causal set theory to construct causal set expressions for certain continuum geometrical quantities. This application of the volume formula is a continuation of work done in~\cite{Buck:2015oaa}. There are many geometric quantities that already have causal set analogues, and in this paper we will add to that list. The more quantities that are accumulated, the more evidence there is that any geometrical quantity can be ``read off" from the causal set. This growing list of quantities also provides evidence for the \emph{Hauptvermutung} --- the conjecture that two very different Lorentzian manifolds cannot be good approximations of the same causal set~\cite{Sorkin:2003bx}.

\section{Volume of a Small Causal Cone}\label{Volume of a Small Causal Cone}

\subsection{The Setup}\label{the setup}

\begin{figure}[t!]
  \centering
    {\includegraphics[scale=0.6]{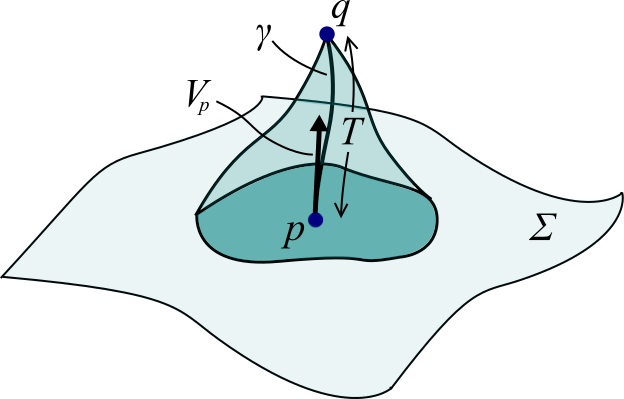}}
     \caption{A illustration of a causal cone in $3$ dimensions of spacetime.}
     \label{fig:causal_cone}
\end{figure}

We will restrict our discussion to a $d$-dimensional, causal, Lorentzian spacetime, $(M,g)$, of finite volume that admits a closed, compact spacelike submanifold, $\Sigma$. A \textit{causal cone} is then constructed in the following way. Choose a \textit{base point} $p\in\Sigma$ and let $\gamma$ be the affinely parameterised geodesic starting at $p$ with tangent vector, $V_p$, normal to $\Sigma$ and future pointing. Travel along this geodesic (in the positive time direction) a proper time $T$, to a point $q$. Past going null rays are sent out from $q$ to form the past light-cone of $q$, denoted by $\partial J^-(q)$. We can then define the \textit{causal cone} to be the region that is the intersection of the future of $\Sigma$ and the past of $q$, i.e. the region $J^+(\Sigma)\cap J^-(q)$. The \textit{base} of the causal cone is the region $\Sigma\cap J^-(q)$ and the upper bounding null surface, the \textit{hat}, is the region $J^+(\Sigma)\cap \partial J^-(q)$. An illustration of this setup is shown in Figure \ref{fig:causal_cone}.

We then ask, what is the spacetime volume of this causal cone as an expansion in small $T$? The terms in front of each power of $T$ in the expansion will be universal, in that they will have the same form for any sufficiently well behaved spacetime. These terms can only depend upon the geometry of the spacetime local to the small causal cone (global topology does not enter the discussion, as we assume the causal cone is small enough to not see it). We can encode this local geometric dependence by having the terms depend upon geometrical quantities evaluated at $p$. If we chose the terms to depend upon geometrical quantities at another point, say $q$, then we could always represent these quantities at $q$ as series expansions in $T$ with coefficients depending upon the quantities evaluated at $p$. In this way one can see that any choice of where to evaluate the geometric quantities (local to the small causal cone) can be related to the choice we make here --- to evaluate them at $p$. We will now introduce all the basic geometric quantities that will arise in the volume formula.

\subsection{Definitions of Geometric Quantities}\label{definitions_of_geometric_objects}

In the next section we will find all the possible geometric objects that can enter into the formula for the volume of the small causal cone, up to the order we are considering. Some of these geometric quantities relate to the past pointing normal vector (one could equally use the future pointing normal) to $\Sigma$, $N_{\Sigma}$, and the future pointing tangent vector along $\gamma$, $V_{\gamma}$. To simplify our search for these quantities in particular we define a vector field that captures the information of both $N_{\Sigma}$ and $V_{\gamma}$. Finding all of the quantities relating to $N_{\Sigma}$ and $V_{\gamma}$ then reduces to enumerating all the possible derivatives of this single vector field, up to the relevant order, and evaluating them at $p$. We now define such a vector field.

Let $x^{\alpha}$ be coordinates for $(M,g$), where $\alpha=0,...,d-1$. Choose a function $S(x)$ (where $x\in M$) that increases to the future, equals zero on $\Sigma$, and equals the proper time along $\gamma$ from $p$ to $x$ for $x\in\gamma$. We then define the covector $n_{\alpha}:=(-g^{\mu\nu} \partial_{\mu}S \partial_{\nu}S)^{-\frac{1}{2}}\partial_{\alpha}S$, and the past pointing vector $n^{\alpha}:=g^{\alpha\beta}n_{\beta}$. When evaluated on the surface, $n^{\alpha}$ are the components of $N_{\Sigma}$, and when evaluated along $\gamma$ they are the components of $-V_{\gamma}$ (the factor of $-1$ comes from the fact that $n^{\alpha}$ is past pointing and the tangent vector to $\gamma$ is future pointing). In this way the vector field $n^{\alpha}$ encodes the vectors $N_{\Sigma}$ and $V_{\gamma}$ at the same time.

The conditions that our chosen function must satisfy afford us a lot of freedom. Any function satisfying the above conditions will give the same vector $n^{\alpha}$ evaluated at $x\in\Sigma\cup\gamma$, but two such functions will in general give rise to vector fields that differ for $x\not\in\Sigma\cup\gamma$. When we choose our function $S(x)$ we are effectively choosing the form of $n^{\alpha}$ away from $\Sigma\cup\gamma$. This choice is independent of our causal cone setup, and any geometric quantities relating to our setup cannot depend on this choice.

We define the induced metric on $\Sigma$ as $h_{\alpha\beta}:=g_{\alpha\beta}+n_{\alpha}n_{\beta}$. If we raise an index with $g^{\alpha\beta}$ then we get the tensor $h^{\alpha}_{\phantom{\alpha}\beta}$ which projects vectors into the tangent space of $\Sigma$, and satisfies $h^{\alpha}_{\phantom{\alpha}\beta}n^{\beta}=0$ and $h^{\alpha}_{\phantom{\alpha}\beta}
h^{\beta}_{\phantom{\beta}\gamma}=h^{\alpha}_{\phantom{\alpha}\gamma}$. The extrinsic curvature tensor is defined as $K_{\alpha\beta}:=n_{\sigma ; \rho}h^{\rho}_{\phantom{\rho}\alpha}
h^{\sigma}_{\phantom{\sigma}\beta}$, where the semi-colon denotes the covariant derivative. The extrinsic curvature scalar is then $K:=K_{\alpha\beta}g^{\alpha\beta}$, and it can be shown that $K=n^{\alpha}_{\phantom{\alpha};\alpha}$ and $K^{\alpha\beta}K_{\alpha\beta}=n^{\alpha}_{\phantom{\alpha};\beta}n^{\beta}_{\phantom{\beta};\alpha}$ on $\Sigma$. The last two relations are both independent of our choice of $n^{\alpha}$ away from $\Sigma$. For more discussion on the geometric quantities mentioned here we refer the reader to~\cite{poisson2004relativist}.

\subsection{All Possible Contributions}\label{All the possible terms}

In this section we will work out a general formula for the volume of a small causal cone, which we denote as $V_\blacktriangle$. In~\cite{Buck:2015oaa} it was found that
\begin{equation}\label{first_order_volume_formula}
V_\blacktriangle(T)=V_{\text{flat}}(T)\bigg(1+\frac{d}{2(d+1)} K T +\mathcal{O}(T^2) \bigg)\;,
\end{equation}
where $K$ is evaluated at $p$, $V_{\text{flat}}(T):=\frac{\text{vol}(S_{d-2})}{d(d-1)}T^d$ is the volume of a flat cone in Minkowski spacetime with a flat base, and $\text{vol}(S_{d-2})$ is the volume of a $(d-2)$-sphere. Here, we are interested in the $\mathcal{O}(T^2)$ term in the brackets in~\eqref{first_order_volume_formula}, which is $\mathcal{O}(T^{d+2})$ if we include the prefactor. The expression multiplying $T^2$ in this term will be a sum of geometric scalars which, by dimensional analysis, must all have dimensions of length $L^{-2}$. The only scalars that contribute are $K^2$, $K^{\alpha\beta}K_{\alpha\beta}$, $R$ and $R_{\alpha\beta}n^{\alpha}n^{\beta}$ (where we have used the usual definitions of the Ricci tensor and Ricci scalar), which we will now show.

To systematically determine all the possible scalar quantities we start with the basic dimensionless objects, $g_{\alpha\beta}$ and $n^{\alpha}$, from which any geometric expression relating to our setup can be constructed. In order to get the right dimensions of length we then form all the scalars involving these objects that contain two derivatives. Every scalar we form will either contain a second order derivative or a product of first order derivatives.

Let us start with the metric $g_{\alpha\beta}$. There are no covariant expressions that can be formed from first order derivatives of $g_{\alpha\beta}$ so we only need to consider its second order derivatives. At second order we have the Riemann tensor, $R_{\alpha\beta\gamma\delta}$, and in order to make a scalar we must contract it with $g^{\alpha\beta}$ and/or $n^{\alpha}$. The only two resulting expressions that can be formed from such contractions are $R$ and $R_{\alpha\beta}n^{\alpha}n^{\beta}$.

There are no terms related to the intrinsic curvature of $\Sigma$ that can be included. For example, the intrinsic Ricci scalar $\,^{d-1}R$ cannot be included as it is not independent from the four quantities we have already, which can be seen from the Gauss-Codazzi equations~\cite{poisson2004relativist}. The only other possibility is the Ricci tensor of $\Sigma$, but we cannot include this as there is nothing to contract it with to give a non-zero quantity other than $\,^{d-1}R$.

We now turn to the vector field $n^{\alpha}$. It still remains to be checked that there are no other scalars that should be included involving a second order derivative or a product of first order derivatives of $n^{\alpha}$. To completely exhaust the latter possibility let us write down the most general (in the sense that we have not contracted any indices) product of two first order derivatives:  $n^{\alpha}_{\phantom{\alpha};\beta}n^{\gamma}_{\phantom{\gamma};\delta}$. We can use the fact that $n^{\alpha}_{\phantom{\alpha};\beta}n^{\beta}=0$ at $p$, and that $n_{\alpha;\beta}n^{\alpha}=\frac{1}{2}(n^{\alpha}n_{\alpha})_{;\beta}=0$ to show that contracting any of the indices with $n^{\mu}$ will give $0$. We, therefore, must contract with the metric to get something non-zero, and one can show that such contractions will give either $K^2$ or $K^{\alpha\beta}K_{\alpha\beta}$. For example, take the following contraction:
\begin{gather}\label{first_order_prod}
\begin{aligned}
n^{\alpha}_{\phantom{\alpha};\beta}n^{\gamma}_{\phantom{\gamma};\delta}g_{\alpha\gamma}g^{\beta\delta}
& =n_{\alpha;\beta}n_{\gamma;\delta}g^{\alpha\gamma}g^{\beta\delta}
\\
& =n_{\alpha;\beta}n_{\gamma;\delta}h^{\alpha\gamma}h^{\beta\delta}
\\
& =n_{\alpha;\beta}n_{\gamma;\delta}
h^{\alpha}_{\phantom{\alpha}\rho}h^{\gamma}_{\phantom{\gamma}\sigma}g^{\rho\sigma}
h^{\beta}_{\phantom{\beta}\mu}h^{\delta}_{\phantom{\delta}\nu}g^{\mu\nu}
\\
& =K_{\mu\rho}K_{\nu\sigma}g^{\rho\sigma}g^{\mu\nu}
\\
& =K^{\nu\sigma}K_{\nu\sigma}\;.
\end{aligned}
\end{gather}
In the first line we have used the fact that $g^{\alpha\beta}=h^{\alpha\beta}-n^{\alpha}n^{\beta}$, and that the resulting contractions with $n^{\mu}$ vanish. In the second line we have used the fact that $h^{\alpha\beta}=h^{\alpha}_{\phantom{\alpha}\gamma}h^{\beta}_{\phantom{\beta}\delta}g^{\gamma\delta}$, and in the third line we have combined the relevant terms to form the two extrinsic curvature tensors. The other possible contractions of $n^{\alpha}_{\phantom{\alpha};\beta}n^{\gamma}_{\phantom{\gamma};\delta}$ with the metric trivially result in either $K^2$ or $K^{\alpha\beta}K_{\alpha\beta}$.

The most general second order derivative of $n^{\alpha}$ is $n^{\alpha}_{\phantom{\alpha};\beta\gamma}$ (in the sense that no indices have been contracted). If we do not contract the bottom two indices with $n^{\beta}n^{\gamma}$ or $h^{\beta}_{\phantom{\beta}\delta}
h^{\gamma}_{\phantom{\gamma}\sigma}$ the resulting expression will depend on our choice of $n^{\alpha}$ away from $\Sigma\cup\gamma$, which cannot be the case for any quantity relating to our geometric setup. To see this let us do the following contraction:  $n^{\alpha}_{\phantom{\alpha};\beta\gamma}n^{\beta}
h^{\gamma}_{\phantom{\gamma}\sigma}$. If we evaluate this at $p$ then the contraction with $n^{\beta}$ projects the first derivative along the geodesic $\gamma$, and the contraction with $h^{\gamma}_{\phantom{\gamma}\sigma}$ projects the second derivative tangent to the surface. Such a second order derivative will depend on the form of $n^{\alpha}$ away from $\Sigma\cup\gamma$. If we stick to contractions with $n^{\beta}n^{\gamma}$ or $h^{\beta}_{\phantom{\beta}\delta}
h^{\gamma}_{\phantom{\gamma}\sigma}$ then we only have to deal with second order derivatives along $\gamma$ or within the surface respectively, which do not depend on $n^{\alpha}$ away from $\Sigma\cup\gamma$. If we contract $n^{\alpha}_{\phantom{\alpha};\beta\gamma}$ with $n^{\beta}n^{\gamma}$ the resulting expression vanishes at $p$, using the fact that $n^{\alpha}_{\phantom{\alpha};\beta}n^{\beta}=0$ along $\gamma$ and that $n^{\alpha}_{\phantom{\alpha};\beta}n_{\alpha}=0$. If we contract with $h^{\beta}_{\phantom{\beta}\delta}
h^{\gamma}_{\phantom{\gamma}\sigma}$ then the two indices $\delta$ and $\sigma$ must be contracted with $g^{\delta\sigma}$, as a contraction with $n^{\delta}$ will give $0$. We also need to contract the free $\alpha$ index, which we can only do with $n_{\alpha}$. The resulting expression does not give anything new, as can be seen by manipulating it as follows:
\begin{gather}
\begin{aligned}
n^{\alpha}_{\phantom{\alpha};\beta\gamma}n_{\alpha}h^{\beta}_{\phantom{\beta}\delta}
h^{\gamma}_{\phantom{\gamma}\sigma}g^{\delta\sigma}
& =n^{\alpha}_{\phantom{\alpha};\beta\gamma}n_{\alpha}g^{\beta\gamma}
\\
& =\left( (n^{\alpha}_{\phantom{\alpha};\beta}n_{\alpha})_{;\gamma} 
- n^{\alpha}_{\phantom{\alpha};\beta}n_{\alpha;\gamma}
\right)g^{\beta\gamma}
\\
& =- n^{\alpha}_{\phantom{\alpha};\beta}n_{\alpha;\gamma}g^{\beta\gamma}
\\
& =- n_{\alpha;\beta}n_{\rho;\gamma}g^{\alpha\rho}g^{\beta\gamma}
\\
&= - K^{\alpha\beta}K_{\alpha\beta}\;.
\end{aligned}
\end{gather}
The equality in the first line comes from the fact that $h^{\beta}_{\phantom{\beta}\delta}
h^{\gamma}_{\phantom{\gamma}\sigma}g^{\delta\sigma}=h^{\beta\gamma}=g^{\beta\gamma}+n^{\beta}n^{\gamma}$, and that the resulting contraction with $n^{\beta}n^{\gamma}$ vanishes, as explained above. The first term in brackets on the second line vanishes as $n^{\alpha}_{\phantom{\alpha};\beta}n_{\alpha}=0$, and in the fourth line we have used~\eqref{first_order_prod}. We have now exhausted the list of possible scalars that can contribute to the volume formula.

The most general formula for the expansion of $V_\blacktriangle(T)$ up to $\mathcal{O}(T^{d+2})$ can be written down as
\begin{gather}\label{general_vol_expansion}
\begin{aligned}
V_\blacktriangle(T)=V_{\text{flat}}(T) & \left(1+\frac{d}{2(d+1)} K T +\left(c_1 K^2 + c_2 K^{\alpha\beta}K_{\alpha\beta} \right.\right.
\\ & \left. +c_3 R+c_4 R_{\alpha\beta}n^{\alpha}n^{\beta}\right)T^2 +\mathcal{O}(T^3) \bigg)\;,
\end{aligned}
\end{gather}
where the geometric quantities are evaluated at $p$. The coefficients $c_1,c_2,c_3$ and $c_4$ can only depend on dimension if they are to be universal. In the next two sections we will use use different spacetime setups to derive these coefficients.

\subsection{Intrinsic Curvature Terms}

In order to determine the coefficients in front of the terms involving $R$ and $R_{\alpha\beta}n^{\alpha}n^{\beta}$ we can follow what was done in~\cite{Gibbons:2007nm}. There, Gibbons and Solodukhin derive the volume formula for a small interval by calculating the volume of an interval in the Einstein static universe and in de Sitter spacetime. We can form causal cones from these two intervals by only considering their ``top-halves".

Specifically, we take the geodesic going from the past-most point of the interval to the future-most point, and take $p$ to be the point half way along that geodesic in proper time. The point $q$ is the future-most point of the interval and the tangent vector of this geodesic at $p$ is $V_p$. This tangent vector is normal to a family of spacelike vectors which generate geodesics expanding out from $p$. We can take the union of these geodesics to be $\Sigma$. Given that we have $p$, $q$ and $\Sigma$, we can construct our causal cone as above. In both spacetimes the base surface generated in this way will have zero extrinsic curvature, and hence we can write the volume expansion as
\begin{equation}\label{general_intrinsic_curve_vol}
V_\blacktriangle(T)=V_{\text{flat}}(t)\left(1+(c_3 R + c_4 R_{\alpha\beta}n^{\alpha}n^{\beta})T^2+\mathcal{O}(T^3)\right)\;.
\end{equation}

In both spacetimes the volume of the causal cone is simply half the volume of interval from which it was constructed. Therefore, the values of the coefficients $c_3$ and $c_4$ can immediately be determined from the interval volume formula in~\cite{Gibbons:2007nm}. We find that
\begin{gather}
\begin{aligned}
& c_3=-\frac{d}{6(d+1)(d+2)}\
\\
& c_4=\frac{d}{6(d+1)}\;.
\end{aligned}
\end{gather}

\subsection{Extrinsic Curvature Terms}

We now move on to the extrinsic curvature terms in \eqref{general_vol_expansion}. For simplicity we can take the spacetime to be Minkowski so that the intrinsic curvature terms all vanish, and look at two curved surfaces within the spacetime. The two surfaces will be specific cases of a one parameter family of surfaces defined by
\begin{equation}
\Sigma_a := \lbrace x\in M | S_a(x)=0 \rbrace\; ,
\end{equation}
where $a$ is the one (positive) parameter and
\begin{equation}
S_a(x):=t-r^2\left( a\cos^2(\theta_1)+\sin^2(\theta_1)\right)\;.
\end{equation}
We have used spherical polar coordinates for the spatial coordinates, such that the coordinates are $x^{\mu}=(t,r,\theta_1,...,\theta_{d-2})$ (where $\theta_{d-2}$ ranges over $[0,2\pi)$ while the others range over $[0,\pi]$). Each surface in the family is not spacelike everywhere, so we chose the base point of the cone, $p$, to be at the origin of the coordinate system where the surfaces are spacelike. We also choose $T$ small enough, with respect to $a$, such that the causal cone's base is a region of the surface that is entirely spacelike. We will first determine the volume of a causal cone for a general choice of the parameter $a$, and then specify at the end to determine the coefficients $c_1$ and $c_2$.

Using the function $S_a(x)$ one can determine the geometric quantities of interest:
\begin{gather}\label{extrinsic_curv_anisotropic_surface}
\begin{aligned}
& K=-2(a+d-2)\;,
\\ & K^{\alpha\beta}K_{\alpha\beta}=4(a^2+d-2)\;,
\end{aligned}
\end{gather}
evaluated at $p$. The volume of the causal cone is
\begin{equation}\label{anisotropic_surface_vol_integral}
\begin{aligned}
V_\blacktriangle(T) & =\int d\Omega_{d-2}\int_0^{r_{int}(\theta_1)}dr\, r^{d-2}\int_{r^2\left(a\cos^2\theta_1+\sin^2\theta_1\right)}^{-r+T}dt
\\
& = \text{vol}(S_{d-3})\int_0^{\pi}d\theta_1\,\sin^{d-3}(\theta_1) \int_0^{r_{int}(\theta_1)}dr\, r^{d-2}\int_{r^2\left(a\cos^2\theta_1+\sin^2\theta_1\right)}^{-r+T}dt
\\
& = \frac{\text{vol}(S_{d-2})}{d(d-1)}T^d \left(1-\frac{d(a+d-2)}{d+1}T \right.
\\
& \hspace{5mm} \left. +\frac{d\left(3a^2+2a(d-2)+d(d-2) \right)}{2(d+1)}T^2+\mathcal{O}(T^3) \right)\;.
\end{aligned}
\end{equation}
In the first line we have evaluated the integrals for the angular coordinates that do not appear in the rest of the integral. In this case the radius of intersection of the hat and base, $r_{int}(\theta_1)$, depends on $\theta_1$. Explicitly this radius is
\begin{equation}\label{r_int_for_anisotropic}
r_{int}(\theta_1)=\frac{-1+\sqrt{1+4aT\cos^2(\theta_1)+4T\sin^2(\theta_1)}}{2\left(a\cos^2(\theta_1)+\sin^2(\theta_1) \right)}\;.
\end{equation}
Given that we only need the $\mathcal{O}(T^2)$ correction to the flat volume of the causal cone, we only require $r_{int}(\theta_1)$ up to $\mathcal{O}(T^2)$. We can, therefore, Taylor expand the RHS of~\eqref{r_int_for_anisotropic} in small $T$ and only keep up to $\mathcal{O}(T^2)$. Using this expansion in place of $r_{int}(\theta_1)$ in~\eqref{anisotropic_surface_vol_integral} makes it possible to evaluate the $\theta_1$ integral and arrive at the final expression.

We equate the $\mathcal{O}(T^2)$ correction in the volume to $c_1 K^2 + c_2 K^{\alpha\beta}K_{\alpha\beta}$, and by using \eqref{extrinsic_curv_anisotropic_surface} we find the following equation for $c_1$ and $c_2$:
\begin{equation}\label{anisotropic_surface_eqn_to_solve}
(a+d-2)^2 c_1 + (a^2+d-2) c_2 = \frac{d\left(3a^2+2a(d-2)+d(d-2) \right)}{8(d+1)}\;.
\end{equation}
We now specialise to the two surfaces given by $a=0$ and $a=1$. Both choices of $a$ can be substituted into~\eqref{anisotropic_surface_eqn_to_solve} to give two simultaneous equations for $c_1$ and $c_2$, the solution for which is
\begin{gather}\label{c1_c2}
\begin{align}
& c_1=\frac{d}{8(d+1)}\;,
\\ & c_2=\frac{d}{4(d+1)}\;.
\end{align}
\end{gather}
The final formula for the volume of a causal cone is then
\begin{gather}\label{final_vol_expansion}
\begin{aligned}
V_\blacktriangle(T)=V_{\text{flat}}(T) & \left(1+\frac{d}{2(d+1)} K T + \frac{d}{4(d+1)}\left(\frac{1}{2}K^2  + K^{\alpha\beta}K_{\alpha\beta} \right.\right.
\\ &\hspace{25mm} \left. -\frac{2}{3(d+2)} R+\frac{2}{3} R_{\alpha\beta}n^{\alpha}n^{\beta}\right)T^2 +\mathcal{O}(T^3) \bigg)\;.
\end{aligned}
\end{gather}
Similar steps can be carried out to determine the formula for the volume of an ``upside down" causal cone, denoted by $V_\blacktriangledown$. Such a cone is constructed by moving backwards in time along the geodesic through $p$ so that $q$ is to the past of $\Sigma$. The absolute value of the proper time between $p$ and $q$ in this case is denoted by $T$. Forward going null rays are then sent out from $q$ till they intersect $\Sigma$. The upside down causal cone is the region $J^-(\Sigma)\cap J^+(q)$ and its volume, $V_\blacktriangledown(T)$, is
\begin{gather}\label{final_vol_expansion_upside_down}
\begin{aligned}
V_\blacktriangledown(T)=V_{\text{flat}}(T) & \left(1-\frac{d}{2(d+1)} K T + \frac{d}{4(d+1)}\left(\frac{1}{2}K^2  + K^{\alpha\beta}K_{\alpha\beta} \right.\right.
\\ &\hspace{25mm} \left. -\frac{2}{3(d+2)} R+\frac{2}{3} R_{\alpha\beta}n^{\alpha}n^{\beta}\right)T^2 +\mathcal{O}(T^3) \bigg)\;.
\end{aligned}
\end{gather}

\section{Use In Causal Set Theory}\label{Use In Causal Set Theory}

In this section we will apply \eqref{final_vol_expansion} and \eqref{final_vol_expansion_upside_down} to causal set theory. This will tie in with what was done in~\cite{Buck:2015oaa}, which one should read for more discussion on this topic. The general motivation for this work is to add to the list of geometric quantities that we can glean from the causal set. Each new quantity added to this list provides further evidence in favor of the Hauptvermutung, and strengthens the idea that the causal set can encode all of spacetime geometry.

\subsection{The Mean of $\mathbf{P}_k$ and $\mathbf{F}_k$}\label{The Mean of $P_k$ and $F_k$}

We say an element of a causal set, $\mathcal{C}$, is of $\mathcal{P}_k$ ($\mathcal{F}_k$) type if it has $k$ elements to its past (future). We define the functions $P_k[\mathcal{C}]$ and $P_k[\mathcal{C}]$ on a causal set, $\mathcal{C}$, to be those that return the number of $\mathcal{P}_k$ and $\mathcal{F}_k$ elements in $\mathcal{C}$ respectively.

We will restrict ourselves to sprinklings into the spacetime $(M,g)$ described above. A sprinkling into such a spacetime naturally generates a partition of the sprinkled causal set, that which has been sprinkled to the future of $\Sigma$, $J^+(\Sigma)$, which we call $\mathcal{C}^+$, and that which has been sprinkled to the past of $\Sigma$, $J^-(\Sigma)$, which we call $\mathcal{C}^-$. We define the random variable $\mathbf{P}_k$ ($\mathbf{F}_k$) as that which takes the value of the function $P_k$ ($F_k$) acting on the sprinkled causal set $\mathcal{C}^+$ ($\mathcal{C}^-$). It should be noted that strictly speaking the random variable is a function of the spacetime, the surface $\Sigma$ and the sprinkling density $\rho$. In this section we aim to find the average over the sprinkling process of $\mathbf{P}_k$ and $\mathbf{F}_k$ as an expansion in large $\rho$. This will allow us to then craft causal set expressions that give continuum geometrical quantities on average in the $\rho\rightarrow\infty$ limit.

The probability that, in a given sprinkling, a point $x\in J^+(\Sigma)$ is a $\mathcal{P}_k$ element of $\mathcal{C}^+$ is given by the probability that $k$ points of the sprinkling lie in the region $J^{-} (x)\cap J^{+} (\Sigma)$. For the Poisson process the probability of such an event is
\be\label{Poisson}
\mathbb P\left (\text{k points in }J^{-} (x)\cap J^{+} (\Sigma)\right)=\frac{\left (\rho\: V_\blacktriangle (x)\right)^k}{k!}e^{-\rho V_\blacktriangle (x)} \;,
\ee
We have written the causal cone volume as a function of the point $x$, as given a point $x$ above $\Sigma$ we can find the geodesic that intersects $x$ with an intital tangent vector normal to $\Sigma$ on the surface. From this geodesic we can get the proper time, $T$, between $\Sigma$ and $x$, which we then insert into the causal cone volume formula in \eqref{final_vol_expansion}. The probability of sprinkling an element into an infinitesimal $d$-volume, $dV_x$, at $x$ is $\rho dV_x$, and so the expected number of $\mathcal{P}_k$ elements above $\Sigma$ is an integral of the product of these probabilities, over all the spacetime points in $J^+(\Sigma)$. We, therefore, have the following expression for the expectation value of $\BP{k}$:
\be\label{eq:nmax}
\left\langle \BP{k}\right\rangle =\rho\int_{J^{+} (\Sigma)}dV_x\; \frac{\left (\rho\: V_\blacktriangle (x)\right)^k}{k!}e^{-\rho V_\blacktriangle (x)} \;.
\ee
Similarly the expected number of $\mathcal{F}_k$ elements below $\Sigma$ is
\be\label{eq:nmin}
\left\langle \BF{k}\right\rangle =\rho\int_{J^{-} (\Sigma)}dV_x\; \frac{\left (\rho\: V_\blacktriangledown (x)\right)^k}{k!}e^{-\rho V_\blacktriangledown (x)} \;,
\ee
where $V_\blacktriangledown (x):= \textrm{vol} (J^{-} (\Sigma)\cap J^{+} (x))$.

In order to evaluate \eqref{eq:nmax} and \eqref{eq:nmin} for large $\rho$ we use ``synchronous'' or Gaussian Normal Coordinates (GNCs) adapted to $\Sigma$ such that in a neighbourhood $U_\Sigma$ of $\Sigma$ the line element is
\be\label{line_element_gncs}
ds^2 = -dt^2 + h_{ij} (t,\mathbf x) dx^i dx^j \;.
\ee
In these coordinates the surface $\Sigma$ corresponds to $t=0$, and the spatial coordinates on $\Sigma$ are $\mathbf{x}$. Each point $x\in U_{\Sigma}$ lies on a unique timelike geodesic with a tangent vector whose components at $\Sigma$ are $-n^\alpha$. The $t$ coordinate of $x$ is equal to the proper time from $\Sigma$ to $x$ along that geodesic. The restriction of the spacetime to this neighbourhood of $\Sigma$ is globally 
hyperbolic with Cauchy surface $\Sigma$.

As $\rho\rightarrow\infty$ we can find a subneighbourhood of $U_\Sigma$ such that the contribution to the integrals from the complement of that subneighbourhood tends to zero exponentially quickly. This is possible because $\Sigma$ is closed and compact, and $J^+(\Sigma)$ and $J^-(\Sigma)$ are of finite volume. More details as to why this is the case are given in~\cite{Buck:2015oaa}. This means that we make only an exponentially small 
error in $\rho$ by cutting off
the integration ranges in \eqref{eq:nmax} and \eqref{eq:nmin} at 
$t=\pm\varepsilon$, with $\varepsilon$ small enough so that we may expand in powers of $t$.

In GNC's we can expand the determinant of the
metric around $t=0$ to write the volume factor as
\begin{gather}\label{volume_form_expansion}
\begin{aligned}
\sqrt{-g} & =h^{\frac{1}{2}}\left(1+
\frac{1}{2}\frac{\dot{h}}{h}t+\frac{1}{4}\left(\frac{\ddot{h}}{h}-\frac{1}{2}\left(\frac{\dot{h}}{h}\right)^2\right)t^2+\mathcal{O}(t^3)\right)
\\
& = h^{\frac{1}{2}}\left(1-Kt+\frac{1}{2}\left(K^2-K^{\alpha\beta}K_{\alpha\beta}-R_{tt}\right)t^2+\mathcal{O}(t^3)\right)
\end{aligned}
\end{gather}
where $h:= det\left (h_{ij}\right)$, and we use a dot for a partial derivative with respect to $t$. All the geometric quantites have been evaluated at $t=0$ and their spatial dependence has been omitted for brevity. To arrive at the second line we have used the fact that, in GNC's, we have that $K=-\dot{h}/2h$ and $\dot{K}=R_{tt}+K^{\alpha\beta}K_{\alpha\beta}$. Using the above expansion, the integrals in~\eqref{eq:nmax} and~\eqref{eq:nmin} become
\begin{gather}\label{eq:nmax_and_eq:nmin}
\begin{aligned}
\left\langle \BF{k}\right\rangle = \rho \int_{\Sigma}d^{d-1}x\int_{-\varepsilon}^{0}dt\:
h^{\frac{1}{2}} & \left(1-Kt+\frac{1}{2}\left(K^2-K^{\alpha\beta}K_{\alpha\beta}-R_{tt}\right)t^2+\mathcal{O}(t^3)\right)
\\
& \times\frac{\left (\rho\: V_\blacktriangledown (t,\mathbf x)\right)^k}{k!} e^{-\rho V_\blacktriangledown (t,\mathbf x)} +\dots \;,
\\
\left\langle \BP{k}\right\rangle = \rho \int_{\Sigma}d^{d-1}x \int_{0}^{\varepsilon}dt\:
h^{\frac{1}{2}} & \left(1-Kt+\frac{1}{2}\left(K^2-K^{\alpha\beta}K_{\alpha\beta}-R_{tt}\right)t^2+\mathcal{O}(t^3)\right)
\\
& \times\frac{\left (\rho\: V_\blacktriangle (t,\mathbf x)\right)^k}{k!} e^{-\rho V_\blacktriangle (t,\mathbf x)} +\dots \;,
\end{aligned}
\end{gather}
where we have used the GNC's and $+ \dots$ denotes  ``terms that vanish exponentially fast with $\rho$ in the limit $\rho \rightarrow \infty$''.

\subsection{Use of the Cone Volume Expansion}\label{section:Use of the Cone Volume Expansion}

We will now use the expansions of the cone volumes \eqref{final_vol_expansion} and \eqref{final_vol_expansion_upside_down} into \eqref{eq:nmax_and_eq:nmin}, and evaluate the integrals in the large $\rho$ limit. We will focus our attention on $\left\langle \BP{k}\right\rangle$ as the case of $\left\langle \BF{k}\right\rangle$ is very similar.

If we substitute in the formula for the volume expansion we get
\begin{gather}\label{sub_in_vol_expansion_first}
\begin{aligned}
\left\langle \BP{k}\right\rangle = \rho  & \int_{\Sigma}d^{d-1}x\, h^{\frac{1}{2}} \int_{0}^{\varepsilon}dt\:
\left(1-Kt+Dt^2+\mathcal{O}(t^3)\right)
\\
& \times\frac{\left (\rho\: At^d\left(1+Bt+Ct^2+\mathcal{O}(t^3)\right)\right)^k}{k!} e^{-\rho At^d\left(1+Bt+Ct^2+\mathcal{O}(t^3)\right)} +\dots \;,
\end{aligned}
\end{gather}
where we have defined
\begin{equation}\label{mean_pk_definitions}
\begin{aligned}
& A := \frac{\text{vol}(S_{d-2})}{d(d-1)}\; ,\;\;\;B:=\frac{d}{2(d+1)} K\; ,
\\
& C:= \frac{d}{4(d+1)}\left(\frac{1}{2} K^2 +K^{\alpha\beta}K_{\alpha\beta} -\frac{2}{3(d+2)} R +\frac{2}{3} R_{tt}\right)\; ,
\\
& D:= \frac{1}{2}\left(K^2-K^{\alpha\beta}K_{\alpha\beta}-R_{tt}\right)\;.
\end{aligned}
\end{equation}
The definitions of $C$ and $D$ are consistent with previous formulae for the volume of the small causal cone as $R_{\alpha\beta}n^{\alpha}n^{\beta}=R_{tt}$ in our setup with the GNC's. Following~\cite{Buck:2015oaa} we will now try and manipulate the integrand into the form of a Gamma function. To do this we split the exponential into a product of two exponentials
\begin{equation}\label{expo_split}
\begin{aligned}
e^{-\rho At^d\left(1+Bt+Ct^2+\mathcal{O}(t^3)\right)} & =e^{-\rho At^d}e^{-\rho At^d\left(Bt+Ct^2+\mathcal{O}(t^3)\right)}
\\
& =e^{-\rho At^d}\left(1-\rho At^d\left(Bt+Ct^2+\mathcal{O}(t^3)\right)+\mathcal{O}\left(t^{2(d+1)}\right) \right)\;.
\end{aligned}
\end{equation}
To get to the second line we have expanded the second exponential on the RHS in the first line in small $t$. If we use~\eqref{expo_split} in~\eqref{sub_in_vol_expansion_first}, and do a small $t$ expansion of the rest of the integrand, then each term in this expansion gives an integral of the following form
\be\label{eq:general_t_n_integral}
\rho^{\eta}\int_{0}^{\varepsilon}dt\
t^{\zeta}e^{-\rho At^{d}} \;,
\ee
where $\eta,\zeta \in \mathbb{R}$. If we make the substitution $z=\rho At^{d}$ then this integral takes on the form of an incomplete gamma function.
\be\label{eq:incomplete_gamma_function}
\frac{A^{-\left (\frac{\zeta+1}{d} \right)}}{d}\rho^{\eta-\left (\frac{\zeta+1}{d} \right)}\int_{0}^{\rho A \varepsilon^d}dz\
z^{\left (\frac{\zeta+1}{d} \right)-1}e^{-z} \;.
\ee
We then take $\rho \rightarrow \infty$ to get
\be\label{eq:gamma_function}
\lim_{\rho\rightarrow\infty}\int_{0}^{\rho A \varepsilon^d}dz\
z^{\left (\frac{\zeta+1}{d} \right)-1}e^{-z}=
\Gamma\left ( \frac{\zeta+1}{d} \right) + \dots \;,
\ee
where, as before, $+ \dots$ denotes terms that tend to zero exponentially fast in $\rho$, and so are zero in the limit $\rho \rightarrow \infty$.

We can now evaluate any integral in~\eqref{sub_in_vol_expansion_first}. This gives us the limiting behaviour of 
$\left\langle \BP{k}\right\rangle$ as 
\begin{gather}\label{eq:nmax_nmin_final}
\begin{aligned}
\left\langle \BP{k}\right\rangle =& \rho^{1-\frac{1}{d}} \frac{A^{-\frac{1}{d}}}{d} \frac{\Gamma\left (\frac{1}{d}+k\right)}{k!}
I_0 - \rho^{1-\frac{2}{d}} \frac{(d+2)A^{-\frac{2}{d}}}{d(d+1)} \frac{\Gamma\left (\frac{2}{d}+k\right)}{k!}
I_1
\\
& + \rho^{1-\frac{3}{d}} \frac{A^{-\frac{3}{d}}}{4 d (d+1)^2 } \frac{\Gamma\left (\frac{3}{d}+k\right)}{k!}
I_2+\mathcal{O}\left(\rho^{1-\frac{4}{d}} \right)\;,
\\
\left\langle \BF{k}\right\rangle =& \rho^{1-\frac{1}{d}} \frac{A^{-\frac{1}{d}}}{d} \frac{\Gamma\left (\frac{1}{d}+k\right)}{k!}
I_0 + \rho^{1-\frac{2}{d}} \frac{(d+2)A^{-\frac{2}{d}}}{d(d+1)} \frac{\Gamma\left (\frac{2}{d}+k\right)}{k!}
I_1
\\
& + \rho^{1-\frac{3}{d}} \frac{A^{-\frac{3}{d}}}{4 d (d+1)^2} \frac{\Gamma\left (\frac{3}{d}+k\right)}{k!}
I_2+\mathcal{O}\left(\rho^{1-\frac{4}{d}} \right)\;.
\end{aligned}
\end{gather}
where we have included $\left\langle \BF{k}\right\rangle$ as well for completeness, and we have defined the integrals over the geometric quantities as
\begin{gather}\label{geom_integrals}
\begin{aligned}
& I_0:=\int_{\Sigma}d^{d-1}x\: \sqrt{h}
\\
& I_1:=\int_{\Sigma}d^{d-1}x\: \sqrt{h}K
\\
& I_2:=\int_{\Sigma}d^{d-1}x\: \sqrt{h} \left(a_1(d) K^2
+a_2(d) K^{\alpha\beta}K_{\alpha\beta}+a_3(d) R + a_4(d) R_{tt}\right)\;,
\end{aligned}
\end{gather}
with
\begin{gather}\label{integral_polynomials}
\begin{aligned}
& a_1(d):=11 + 2 d (d+5)
\\
& a_2(d):=-(d+1)(2 d+5)
\\
& a_3(d):=\frac{2(d+1)}{(d+2)}
\\
& a_4(d):=-2(d+1)(d+2)\;.
\end{aligned}
\end{gather}

The Gauss-Codazzi equations relate the four geometric quantities in $I_2$ to the Ricci scalar for the surface, $\,^{d-1}R$. We can use this relation to swap out any of the four quantities in $I_2$ for $\,^{d-1}R$.

\subsection{Causal Set Expressions}\label{a_causal_set_expression}

With~\eqref{eq:nmax_nmin_final} in hand we can form causal set expressions to give us any of the integrals in~\eqref{geom_integrals}. In~\cite{Buck:2015oaa} causal set expressions were constructed for $I_0$ and $I_1$, so we will focus on an expression for $I_2$. These expressions are constructed such that their average over the sprinkling process tends to the desired integral in the $\rho\rightarrow\infty$ limit.

Take the following causal set function
\begin{equation}\label{general_causet_function}
I\left[\mathcal{C},\mathcal{C}^+,\mathcal{C}^- ;\vec{p},\vec{q}\right]:=l^{d-3}A_d\left( \sum_m p_m \F{m}\left[\mathcal{C}^- \right]
+  \sum_n q_n \P{n}\left[\mathcal{C}^+ \right] \right)\;,
\end{equation}
where $l=\rho^{-\frac{1}{d}}$ is the discreteness length and $A_d$ is a real constant that depends only on dimension. $\vec{p}$ and $\vec{q}$ denote strings of real constants $p_m$ and $q_n$. The sums run over the non-negative integers, although they may be terminated if the constants $p_m$ and $q_n$ are $0$ for $m$ or $n$ larger than some integer. We have also partitioned the causal set into $\mathcal{C}^+$ and $\mathcal{C}^-$, and restricted the functions $F_m$ and $P_n$ to act on $\mathcal{C}^-$ and $\mathcal{C}^+$ respectively. We can think of this as a family of functions, with each member of the family specified by their particular strings $\vec{p}$ and $\vec{q}$. The strings are not totally arbitrary, however, and soon we will see that they must satisfy certain constraints if we are to recover $I_2$.

We define the random variable $\mathbf{I}$ as that which takes the value $I\left[\mathcal{C},\mathcal{C}^+,\mathcal{C}^- ;\vec{p},\vec{q}\right]$ under sprinkling into $M$. This random variable depends on the spacetime $M$, surface $\Sigma$, density $\rho$, and the strings $\vec{p}$ and $\vec{q}$.

$\mathbf{I}$ can be written in terms of the random variables $\mathbf{F}_m$ and $\mathbf{P}_n$, where we recall that these random variables realise the values of $\F{m}\left[\mathcal{C}^- \right]$ and $\P{n}\left[\mathcal{C}^+ \right]$ under a sprinkling into $M$. Writing $\mathbf{I}$ in this way gives
\begin{equation}\label{general_causet_random_variable}
\mathbf{I}:=\rho^{\frac{3}{d}-1} A_d\left( \sum_m p_m \mathbf{F}_m
+  \sum_n q_n \mathbf{P}_n \right)\;,
\end{equation}
where we have omitted the arguments of the random variables for brevity.
We want to take the expectation value of this random variable over the sprinkling process and extract out the integral $I_2$ in the limit of $\rho\rightarrow\infty$. That is, we are aiming for
\begin{equation}\label{mean_of_i_to_get_integral}
\lim_{\rho\rightarrow\infty} \left\langle \mathbf{I} \right\rangle=I_2\;.
\end{equation}

Given we have~\eqref{eq:nmax_nmin_final} we can take the expectation value of~\eqref{general_causet_random_variable} in the limit of $\rho\rightarrow\infty$ to find
\begin{gather}\label{eq:boundary_final_proof}
\begin{aligned}
\lim_{\rho\rightarrow\infty} & \left\langle \textbf{I}\right\rangle = \lim_{\rho\rightarrow\infty} \rho^{\frac{3}{d}-1} A_{d}\left (\sum_m p_m \left\langle\textbf{F}_m\right\rangle  + \sum_n q_n \left\langle\textbf{P}_n\right\rangle\right)
\\
= & A_d\lim_{\rho\rightarrow\infty}\Bigg[\rho^{\frac{2}{d}}\frac{A^{-\frac{1}{d}}}{d}\left(\sum_m p_m \frac{\Gamma\left (\frac{1}{d}+m \right)}{m!}  + \sum_n q_n\frac{\Gamma\left (\frac{1}{d}+n \right)}{n!} \right) I_0
\\
+ &\rho^{\frac{1}{d}}\frac{(d+2)A^{-\frac{2}{d}}}{d(d+1)}\left(\sum_m\frac{\Gamma\left (\frac{2}{d}+m \right)}{m!}  - \sum_n q_n\frac{\Gamma\left (\frac{2}{d}+n \right)}{n!} \right) I_1
\\
+ &\frac{A^{-\frac{3}{d}}}{4d(d+1)^2}\left(\sum_m\frac{\Gamma\left (\frac{3}{d}+m \right)}{m!}  + \sum_n q_n\frac{\Gamma\left (\frac{3}{d}+n \right)}{n!} \right) I_2 \Bigg]\;.
\\
\end{aligned}
\end{gather}
In order for~\eqref{mean_of_i_to_get_integral} to be satisfied we get the following conditions on $\vec{p}$ and $\vec{q}$:
\begin{gather}\label{coefficient_relation}
\begin{aligned}
& \sum_m p_m \frac{\Gamma\left (\frac{1}{d}+m \right)}{m!}  + \sum_n q_n\frac{\Gamma\left (\frac{1}{d}+n \right)}{n!}=0 \;,
\\
& \sum_m p_m \frac{\Gamma\left (\frac{2}{d}+m \right)}{m!}  - \sum_n q_n\frac{\Gamma\left (\frac{2}{d}+n \right)}{n!}=0 \;,
\\
& \sum_m p_m \frac{\Gamma\left (\frac{3}{d}+m \right)}{m!}  + \sum_n q_n\frac{\Gamma\left (\frac{3}{d}+n \right)}{n!}=1  \;.
\end{aligned}
\end{gather}
We also find that the constant $A_d$ must be
\begin{equation}\label{constant_ad}
A_d=4d(d+1)^2A^{\frac{3}{d}}\;.
\end{equation}
There are many different $\vec{p}$ and $\vec{q}$ strings that satisfy~\eqref{coefficient_relation}. This freedom comes from the fact that we are effectively discretising a mix of second order derivatives. From~\eqref{coefficient_relation} we see that at least three non-zero entries in $\vec{p}$ and $\vec{q}$ are needed, which is consistent with the idea that it is a discrete second order derivative.

The simplest causal set expressions that give $I_2$ (in the sense of~\eqref{mean_of_i_to_get_integral}) are those formed by taking only the smallest $k$ components of $p_k$ and $q_k$ to be non-zero. For example, if the only non-zero components are $p_0$, $q_0$ and $q_1$ then solving~\eqref{coefficient_relation} gives
\begin{equation}\label{pk_qk_simple}
p_0=\frac{1}{4 \Gamma\left(\frac{3}{d}\right)}\;,\;\; q_0=-\frac{3}{4 \Gamma\left(\frac{3}{d}\right)}\;,\;\;q_1=\frac{d}{2 \Gamma\left(\frac{3}{d}\right)}\;,
\end{equation}
with all other components equal to $0$.

We denote the strings with these as the only non-zero components by $\vec{p}_a$ and $\vec{q}_a$. Inserting these strings into~\eqref{general_causet_function} simplifies the causal set function to
\begin{equation}\label{causet_function_simple}
I\left[\mathcal{C},\mathcal{C}^+,\mathcal{C}^- ;\vec{p}_a,\vec{q}_a\right]=l^{d-3}\frac{A_d}{4\Gamma\left(\frac{3}{d}\right)}\left(\F{0}\left[\mathcal{C}^- \right]-3\P{0}\left[\mathcal{C}^+ \right]
+  2d \P{1}\left[\mathcal{C}^+ \right] \right)\;
\end{equation}
We define $I_a[\mathcal{C},\mathcal{C}^+,\mathcal{C}^-]$ as the function on the RHS in~\eqref{causet_function_simple}, and its random variable counterpart as $\textbf{I}_a$, where the counterpart is formed in the usual way by promoting the functions $F_m$ and $P_n$ to random variables.

One can also take an entire string to be zero. For example, take $\vec{p}$ to be the zero string $\vec{0}$ (every component $p_k=0$). If we take the first $3$ components of $\vec{q}$ to be the only non-zero components then, by solving~\eqref{coefficient_relation}, we find
\begin{equation}\label{qk_simple}
q_0=\frac{1}{\Gamma\left(\frac{3}{d}\right)}\;,\;\; q_1=-\frac{d(d+3)}{2 \Gamma\left(\frac{3}{d}\right)}\;,\;\;q_2=\frac{d^2}{\Gamma\left(\frac{3}{d}\right)}\;.
\end{equation}
We denote the string with these as the only non-zero components as $\vec{q}_-$.

If these strings are inserted into the arguments of $I$ we find
\begin{equation}\label{causet_function_simple_q_only}
I\left[\mathcal{C},\mathcal{C}^+,\mathcal{C}^- ;\vec{0},\vec{q}_-\right]=l^{d-3}\frac{A_d}{\Gamma\left(\frac{3}{d}\right)}\left(\P{0}\left[\mathcal{C}^+ \right]-\frac{d(d+3)}{2}\P{1}\left[\mathcal{C}^+ \right] +d^2 \P{2}\left[\mathcal{C}^+ \right]\right)\;.
\end{equation}
The causal set $\mathcal{C}^-$ does not enter on the RHS as there are no $F_k$ functions to act on it, and hence we can view the RHS as being a function on a single causal set, $\mathcal{C}^+$, without reference to it being part of some larger causal set. We define the function $I_-[\mathcal{C}^+]$ as the RHS of~\eqref{causet_function_simple_q_only}. $I_-$ is really a function of a single causal set, as it does not depend on $\mathcal{C}^-$. The random variable counterpart, $\mathbf{I}_-$, can be formed in the usual way. This random variable does not depend on $J^-(\Sigma)$, and so can be viewed as being a function of the spacetime $J^+(\Sigma)$ and its past boundary $\Sigma$ only (and the density $\rho$). Thus, given a single causal set, $\mathcal{C}$, we can think of $I_-[\mathcal{C}]$ as the causal set analogue of the geometrical quantity $I_2$  corresponding, in some sense, to the ``past boundary" of $\mathcal{C}$. A similar expression can be formed for the ``future boundary" of a causal set by taking $\vec{q}=\vec{0}$ and having only the first $3$ components of $\vec{p}$ be non-zero.

The continuum quantity, $I_2$, for which we have constructed the family of causal set expressions, $I$, contains four different geometric quantities. This means that given some causal set, and the value of $I$ acting on that causal set, we do not know what contribution to that number has come from the causal set analogue of one of the four geometric quantities in $I_2$ alone. We also do not know whether this value will be close to the continuum value of some manifold from which our causal set could arise as a typical sprinkling. This question will be addressed in the next section.

The family of causal set functions found here are not as immediately useful as causal set functions that correspond to a single geometrical quantity, such as those found in~\cite{Buck:2015oaa} for the integrals $I_0$ and $I_1$. We can attempt to extract a single quantity from the integral $I_2$ using other causal set expressions alongside $I$. We will now sketch out how one might attempt to extract $K^{ab}K_{ab}$. The number of causal set elements in an interval gives the spacetime volume of the interval, and using the formulae in~\cite{Myrheim:1978,Gibbons:2007nm} one might be able to extract $c_3 R + c_4 R_{\alpha\beta}n^{\alpha}n^{\beta}$ from this number. One could also determine $K$ from the causal set expressions given in~\cite{Buck:2015oaa}~\footnote{The causal set expressions in~\cite{Buck:2015oaa} give the spatial volume of the hypersurface and the extrinsic curvature integrated over the hypersurface. If we divide the latter by the former we get the average value of the extrinsic curvature over the hypersurface, $K_{avg}$. We can only use the expressions in~\cite{Buck:2015oaa} to determine $K$ when the region of the hypersurface that we are interested in is such that $K_{avg}\approx K$ for any point in that region.}. The remaining quantity in $I_2$, $K^{\alpha\beta}K_{\alpha\beta}$, can then be extracted on its own.

\subsection{Fluctuations}

So far we have studied the expectation value of the random variable in~\eqref{general_causet_random_variable}. In this section we turn towards the fluctuations about this average.

We would like a typical realisation of the random variable $\textbf{I}$ on a single causal 
set, sprinkled into $M$ with density $\rho$, to be close to the continuum geometrical quantity $I_2$. For this we require that (i) the mean value be close to $I_2$ at this sprinkling density, and that (ii) the fluctuations about the mean are small.

For (i) to be satisfied we need the approximations and expansions in the previous sections to be valid. This will be the case when the curvature scales involved in the setup are much larger than the discreteness length $l=\rho^{-\frac{1}{d}}$, so that the sprinkling is dense enough to encode the geometric information of the spacetime. For more discussion on this point we refer the reader to~\cite{Buck:2015oaa}.

We now ask when (ii) will be satisfied. To estimate how the fluctuations of $\textbf{I}$, or more specifically the standard deviation, $\sigma[\textbf{I}]=\text{Var}[\textbf{I}]^\frac12$, depends on $\rho$ we can use the same heuristic argument as in~\cite{Buck:2015oaa}. The entire argument does not hold in this case, but the numerics seem to support it, and hence we describe it here anyway.

Take any spacetime region of fixed volume $V$. The number of causal set elements in a sprinkling is a random variable, $\textbf{N}$, with mean $\left\langle\textbf{N}\right\rangle=\rho V$ and standard deviation $\sqrt{\left\langle\textbf{N}\right\rangle}$, as the sprinkling is a Poisson process. If one imagines thickening the hypersurface $\Sigma$ by one unit of the discreteness scale $l=\rho^{-\frac{1}{d}}$ (by Lie dragging the surface along its normal by an amount $l$), then the volume of this thickened surface is approximately $\mathrm{vol}(\Sigma) l=\mathrm{vol}(\Sigma)\rho^{-\frac{1}{d}}$. $\BF{k}$ and $\BP{k}$ are random variables that, in some sense, count nearest neighbours of $\Sigma$. We, therefore, expect their mean values to scale like $\rho\mathrm{vol}(\Sigma)\l=\mathrm{vol}(\Sigma)\rho^\frac{d-1}{d}\propto\left\langle\textbf{N}\right\rangle^\frac{d-1}{d}$. This would suggest that $\mathbf P_k$ and $\mathbf F_k$ will
have standard deviations of order $\left\langle\textbf{N}\right\rangle^\frac{d-1}{2d} = (\rho V)^\frac{d-1}{2d}$. In~\cite{Buck:2015oaa} the argument continues by using the independence of the random variables $\BF{0}$ and $\BP{0}$ that enter into the boundary term in question. In our case, however, every member of the family of random variables in~\eqref{general_causet_random_variable} will contain at least two random variables that are not independent. This is due to the fact that every member of the family must have at least two $\BF{k}$'s or two $\BP{k}$'s (examples of this can be seen above in~\eqref{causet_function_simple} and~\eqref{causet_function_simple_q_only}). In~\cite{Buck:2015oaa} it was found that the fluctuations supported this heuristic argument even in the case where they formed a boundary term with random variables that were not independent. We shall, therefore, proceed with the next step in the argument in the hope that it will be supported by numerical evidence. The next step in our case is to say that the standard deviation of $\textbf{I}$ will go like that of $\BF{k}$ or $\BP{k}$ (as $\rho^\frac{d-1}{2d}$) but multiplied by the dependece on $\rho$ from the factor of $l^{d-3}$ (or $\rho^{\frac{3-d}{d}}$ in terms of the density) at the front of the RHS in~\eqref{general_causet_random_variable}. That is, $\sigma[\textbf{I}]$ should scale like $\rho^{\frac{3-d}{d}} \rho^{\frac{d-1}{2d}}=\rho^{\frac{5-d}{2d}}$, or as $\left\langle\textbf{N}\right\rangle^{\frac{5-d}{2d}}$ in terms of the mean number of elements sprinkled.
\begin{figure}[t]
  \centering
    {\includegraphics[scale=0.4]{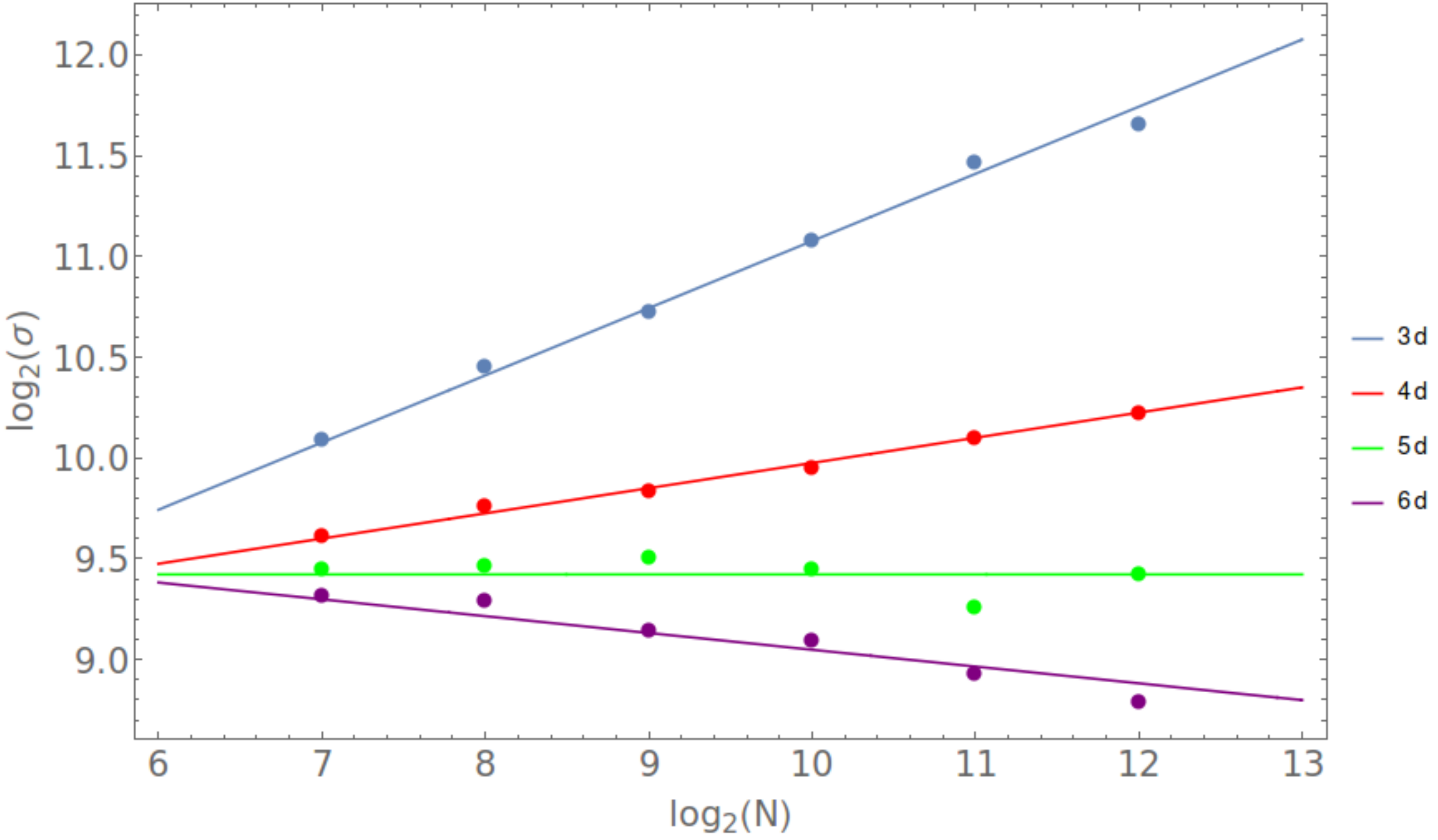}}
     \caption{Base-$2$ log-log plot of the standard deviation of $\textbf{I}_a$ against $\left\langle\textbf{N}\right\rangle$. In the graph these quantities have been denoted by $\sigma$ and $N$ respectively. From top to bottom the data and the corresponding best fit lines are for dimensions $d=3,4,5$ and $6$ respectively.}
     \label{fig:fluctuations}
\end{figure}
\FloatBarrier
This scaling law was tested numerically by sprinkling into a $d$-dimensional cube, $[0,1]^d$, in Minkowski space of that dimension. $\Sigma$ was taken to be the surface $t=1/2$. The mean number of sprinkled elements, $\left\langle\textbf{N}\right\rangle$, ranged from $2^7$ to $2^{12}$, and for each mean number of elements we did $400$ sprinklings. The mean and standard deviation of $\textbf{I}_a$ was then calculated for the sample of $400$ sprinklings at each $\left\langle\textbf{N}\right\rangle$. In this setup the mean of $\textbf{I}_a$ is zero as the surface and spacetime are flat. This was done for dimensions $d=3,4,5,6$ and the results of $\log_2(\sigma[\textbf{I}_a])$ against $\log_2(\left\langle\textbf{N}\right\rangle)$ can be seen in Figure~\ref{fig:fluctuations}.
The fitted lines have the form $\frac{5-d}{2d}\log_2(\left\langle\textbf{N}\right\rangle)+\xi$, where the constants $\xi$ for each dimension are of order $1$. The results for $\sigma[\textbf{I}_-]$ also show a similar scaling.

These results suggest that the heuristic argument is correct as it has predicted the right scaling. Unfortunately, this means that in $4$ dimensions the causal set random variable $\textbf{I}$ has fluctuations that grow with $\left\langle\textbf{N}\right\rangle$, much like the Benincasa-Dowker-Glaser action~\cite{Benincasa:2010ac,Sorkin:2007qi,Dowker:2013vba,Belenchia:2015hca}. In the case of the action one can modify it with an intermediate length scale to dampen the fluctuations. Perhaps this can be done here too. Witout this damping we can only satisfy (ii) for large $\rho$ if $d>5$. Further work should be done to determine if the scaling of the fluctuations persists in cases where the spacetime and/or the hypersurface are not flat.

\subsection{Normal Derivatives of a Scalar Field}

The techniques in Section \ref{a_causal_set_expression} can be used to find causal set expressions relating to the normal derivatives of a scalar field. A scalar field on a causal set is a function from the causal set to the real numbers (or complex numbers for a complex scalar field). These real numbers can be denoted by $\phi_i$ where the index $i$ runs over the causal set elements.

The functions $F_k$ and $P_k$ sum up the number of $\mathcal{F}_k$ and $\mathcal{P}_k$ elements respectively. We will now define functions that sum up the values of $\phi_i$ on those particular elements. Explicitly, we define these new functions as
\begin{equation}\label{f_phi_def}
F^{\phi}_k[\mathcal{C}]=\sum_{i\in\lbrace\mathcal{F}_k
\rbrace} \phi_i\;,
\end{equation}
\begin{equation}\label{p_phi_def}
P^{\phi}_k[\mathcal{C}]:=\sum_{i\in\lbrace\mathcal{P}_k
\rbrace} \phi_i\;,
\end{equation}
where we take $\lbrace\mathcal{F}_k\rbrace$ and $\lbrace\mathcal{P}_k\rbrace$ to denote lists of the indices of the $\mathcal{F}_k$ and $\mathcal{P}_k$ elements respectively, so that the sums run over these elements. These functions depend on the causal set and the scalar field values, $\phi_i$, defined on that causal set.

Under the sprinkling process the scalar field on the causal set defines random variables for the functions in~\eqref{f_phi_def} and~\eqref{p_phi_def} in the following way. We start with the usual notion of a  scalar field, $\phi(x)$, defined on the manifold, $M$, we wish to sprinkle in to. The sprinkling generates a random causal set, $\mathcal{C}$, and we take $x_i$ to be the spacetime point of the $i$th element. The scalar field value on the $i$th element of the causal set is then simply the value of the scalar field $\phi(x)$ evaluated at $x_i$, ie. $\phi_i=\phi(x_i)$. At this point we could define random variables to return the values of $F^{\phi}_k[\mathcal{C}]$ and $P^{\phi}_k[\mathcal{C}]$ under the sprinkling process. We will instead follow in the footsteps of Section~\ref{a_causal_set_expression} and define the random variables $\mathbf{F}^{\phi}_k$ and $\mathbf{P}^{\phi}_k$ as those that return the values of the functions $F^{\phi}_k[\mathcal{C}^-]$ and $P^{\phi}_k[\mathcal{C}^+]$ respectively, where we have the same spacetime setup as before. These random variables, for a given $k$, are functions of the manifold, sprinkling density, surface $\Sigma$, and the scalar field $\phi(x)$.

We wish to find the expectation values of theses random variables in the hope that we can use them to construct causal set expressions for continuum quantities. To get the expectation value of $\mathbf{P}^{\phi}_k$, say, we need to take the product of the probability for an element to have been sprinkled in an infinitesimal volume element at $x$, times the probability that it is a $\mathcal{P}_k$ element, times the value of the scalar field at $x$, $\phi(x)$, and then integrated over all $x$ in the region to the future of $\Sigma$. The expectation value is then
\be\label{eq:p_phi_general_expectation_val}
\left\langle \mathbf{P}^{\phi}_k\right\rangle =\rho\int_{J^{+} (\Sigma)}dV_x\; \phi(x)\frac{\left (\rho\: V_\blacktriangle (x)\right)^k}{k!}e^{-\rho V_\blacktriangle (x)} \;.
\ee
Likewise, for $\mathbf{F}^{\phi}_k$ we have
\be\label{eq:f_phi_general_expectation_val}
\left\langle \mathbf{F}^{\phi}_k\right\rangle =\rho\int_{J^{-} (\Sigma)}dV_x\; \phi(x)\frac{\left (\rho\: V_\blacktriangledown (x)\right)^k}{k!}e^{-\rho V_\blacktriangledown (x)} \;.
\ee

We can, once again use GNCs, $x^\mu= (t,\mathbf x)$, adapted to $\Sigma$ such that in a neighbourhood $U_\Sigma$ of $\Sigma$ the line element is given by~\eqref{line_element_gncs}, and $\Sigma$ is the surface defined by $t=0$. The integrals can be simplified as before, so that we only integrate to $\varepsilon$ in the time coordinate, and only make an exponentially small error in doing so. The addition of $\phi(x)$ in the integrand will not change this fact as it does not depend on $\rho$ and so will not alter how the integrand changes with $\rho$. We take $\rho$ to be large enough such that $\varepsilon$ can be small enough for us to expand the scalar field and the determinant of the spatial metric $h$ about $\Sigma$. We can now write the expectation values as
\begin{gather}\label{phi_expansions_of_integrals}
\begin{aligned}
\left\langle \mathbf{F}^{\phi}_k\right\rangle = & \rho \int_{\Sigma}d^{d-1}x\int_{-\varepsilon}^{0}dt\:
h^{\frac{1}{2}} \left(1-Kt+\frac{1}{2}\left(K^2-K^{\alpha\beta}K_{\alpha\beta}-R_{tt}\right)t^2+\mathcal{O}(t^3)\right)
\\
& \times\left(\phi+\dot{\phi}t+\frac{1}{2}\ddot{\phi}t^2+\mathcal{O}(t^3) \right)\frac{\left (\rho\: V_\blacktriangledown (t,\mathbf x)\right)^k}{k!} e^{-\rho V_\blacktriangledown (t,\mathbf x)} +\dots \;,
\\
\left\langle \mathbf{P}^{\phi}_k\right\rangle = & \rho \int_{\Sigma}d^{d-1}x \int_{0}^{\varepsilon}dt\:
h^{\frac{1}{2}}\left(1-Kt+\frac{1}{2}\left(K^2-K^{\alpha\beta}K_{\alpha\beta}-R_{tt}\right)t^2+\mathcal{O}(t^3)\right)
\\
& \times\left(\phi+\dot{\phi}t+\frac{1}{2}\ddot{\phi}t^2+\mathcal{O}(t^3) \right)\frac{\left (\rho\: V_\blacktriangle (t,\mathbf x)\right)^k}{k!} e^{-\rho V_\blacktriangle (t,\mathbf x)} +\dots \;,
\end{aligned}
\end{gather}
where the dots above $\phi$ denote time derivatives and all of the geometric quantities are defined similarly to~\eqref{volume_form_expansion}. The terms $\phi$, $\dot{\phi}$ and $\ddot{\phi}$ are evaluated at $t=0$ and depend on the surface coordinate $\mathbf x$. Again, we use $+ \dots$ to stand for ``terms that vanish exponentially fast in the limit $\rho \rightarrow \infty$''.

We follow the same procedure as in Section~\ref{section:Use of the Cone Volume Expansion} to expand the cone volumes in $t$ and evaluate the integrals by transforming them into the form of Gamma functions. The only difference here is that one must take into account of one more expansion, that of the scalar field. Because of the similarities we will just state the final expansion in large $\rho$ for both of the required expectation values:
\begin{gather}\label{eq:phi_expectation_final}
\begin{aligned}
\left\langle \mathbf{P}_k^{\phi}\right\rangle =& \rho^{1-\frac{1}{d}} \frac{A^{-\frac{1}{d}}}{d} \frac{\Gamma\left (\frac{1}{d}+k\right)}{k!}
I_0^{\phi} - \rho^{1-\frac{2}{d}} \frac{A^{-\frac{2}{d}}}{d} \frac{\Gamma\left (\frac{2}{d}+k\right)}{k!}\left(
\frac{(d+2)}{(d+1)}I_1^{\phi}-I_0^{\dot{\phi}}\right)
\\
+ \rho^{1-\frac{3}{d}} & \frac{A^{-\frac{3}{d}}}{d} \frac{\Gamma\left (\frac{3}{d}+k\right)}{k!}
\left(\frac{1}{4(d+1)^2}I_2^{\phi}-
\frac{(2d+5)}{2(d+1)}I_1^{\dot{\phi}}+
\frac{1}{2}I_0^{\ddot{\phi}}\right)
+\mathcal{O}\left(\rho^{1-\frac{4}{d}} \right)\;,
\\
\left\langle \mathbf{F}_k^{\phi}\right\rangle =& \rho^{1-\frac{1}{d}} \frac{A^{-\frac{1}{d}}}{d} \frac{\Gamma\left (\frac{1}{d}+k\right)}{k!}
I_0^{\phi} + \rho^{1-\frac{2}{d}} \frac{A^{-\frac{2}{d}}}{d} \frac{\Gamma\left (\frac{2}{d}+k\right)}{k!}\left(
\frac{(d+2)}{(d+1)}I_1^{\phi}-I_0^{\dot{\phi}}\right)
\\
+ \rho^{1-\frac{3}{d}} & \frac{A^{-\frac{3}{d}}}{d} \frac{\Gamma\left (\frac{3}{d}+k\right)}{k!}
\left(\frac{1}{4(d+1)^2}I_2^{\phi}-
\frac{(2d+5)}{2(d+1)}I_1^{\dot{\phi}}+
\frac{1}{2}I_0^{\ddot{\phi}}\right)
+\mathcal{O}\left(\rho^{1-\frac{4}{d}} \right)\;,
\end{aligned}
\end{gather}
where we have added superscripts to the integrals $I_{0,1,2}$ to mean that one must include whatever is in the superscript in the integrand. For example,
\begin{equation}\label{example_with_superscript}
I_1^{\dot{\phi}}=\int_{\Sigma}d^{d-1}x\: \sqrt{h}K \dot{\phi}\;,
\end{equation}
which is the integral $I_1$ with the integrand multiplied by $\dot{\phi}$.

We can now define causal set expressions utilising~\eqref{eq:phi_expectation_final} that give the different integrals in the expansion. First, we will construct an expression for $I_0^{\phi}$. We define
\begin{equation}\label{general_causet_function_integral_phi}
J_0^{\phi}\left[\mathcal{C},\mathcal{C}^+,\mathcal{C}^- ;\vec{p},\vec{q}\right]:=l^{d-1}d A^{\frac{1}{d}}\left( \sum_m p_m F_m^{\phi}\left[\mathcal{C}^- \right]
+  \sum_n q_n P_n^{\phi}\left[\mathcal{C}^+ \right] \right)\;,
\end{equation}
where $p_m$ and $q_n$ are strings of real numbers that satisfy
\begin{equation}\label{coefficient_relation_integral_phi}
\sum_m p_m \frac{\Gamma\left (\frac{1}{d}+m \right)}{m!}  + \sum_n q_n\frac{\Gamma\left (\frac{1}{d}+n \right)}{n!}=1 \;.
\end{equation}
Only one of the components needs to be non-zero to satisfy~\eqref{coefficient_relation_integral_phi}.

We can also define the random variable counterpart, $\mathbf{J}_0^{\phi}$, in the usual way, by promoting $F^{\phi}_k$ and $P^{\phi}_k$ to random variables. Given that the coefficients satisfy~\eqref{coefficient_relation_integral_phi} one can follow the same steps as in~\eqref{eq:boundary_final_proof} to show that
\begin{equation}\label{mean_of_j0_to_get_integral}
\lim_{\rho\rightarrow\infty} \left\langle \mathbf{J}_0^{\phi} \right\rangle= I_0^{\phi}\;,
\end{equation}
where we have omitted the arguments of $\mathbf{J}_0^{\phi}$, which are the spacetime $M$, the surface $\Sigma$, the density $\rho$, the field $\phi(x)$, and the strings $\vec{p}$ and $\vec{q}$. 

The simplest choices of $p_m$ and $q_n$ are those in which there is only one non-zero component. If we take the first element of $\vec{p}$ to be the only non-zero one, so that $\vec{q}=\vec{0}$, then $p_0=\Gamma(\frac{1}{d})^{-1}$ solves~\eqref{coefficient_relation_integral_phi}. Using these strings the RHS of~\eqref{general_causet_function_integral_phi} becomes
\begin{equation}
l^{d-1}\frac{dA^{\frac{1}{d}}}{\Gamma\left(\frac{1}{d} \right)}F_0^{\phi}\left[ \mathcal{C}^-\right]\;.
\end{equation}
For the opposite case where $\vec{p}=\vec{0}$ and $q_0$ is the only non-zero component we get a causal set function which is proportional to $P_0^{\phi}\left[\mathcal{C}^+ \right]$. These two causal set functions have corresponding random variables whose expectation values give $\int_{\Sigma}d^{d-1}x\sqrt{h}\phi$ in the $\rho\rightarrow\infty$ limit. This seems intuitively correct, as one would expect that summing the values of the scalar field at the causal set elements close to the surface will give something like $\int_{\Sigma}d^{d-1}x\sqrt{h}\phi$ in the continuum limit.

Next, we would like construct a causal set expression for the part in brackets in the second term on the RHS in~\eqref{eq:phi_expectation_final}. We define the causal set function
\begin{equation}\label{general_causet_function_integral_phi_dot}
J_1^{\phi}\left[\mathcal{C},\mathcal{C}^+,\mathcal{C}^- ;\vec{p},\vec{q}\right]:=l^{d-2}d A^{\frac{2}{d}}\left( \sum_m p_m F_m^{\phi}\left[\mathcal{C}^- \right]
+  \sum_n q_n P_n^{\phi}\left[\mathcal{C}^+ \right] \right)
\end{equation}
where the strings $p_m$ and $q_n$ now satisfy
\begin{gather}\label{coefficient_relation_integral_phi_dot}
\begin{aligned}
& \sum_m p_m \frac{\Gamma\left (\frac{1}{d}+m \right)}{m!}  + \sum_n q_n\frac{\Gamma\left (\frac{1}{d}+n \right)}{n!}=0 \;,
\\
& \sum_m p_m \frac{\Gamma\left (\frac{2}{d}+m \right)}{m!}  - \sum_n q_n\frac{\Gamma\left (\frac{2}{d}+n \right)}{n!}=1 \;.
\end{aligned}
\end{gather}
With this, one can verify that the random variable counterpart, $\mathbf{J}_1^{\phi}$, for~\eqref{general_causet_function_integral_phi_dot} satisfies
\begin{equation}\label{mean_of_j1_to_get_integral}
\lim_{\rho\rightarrow\infty} \left\langle \mathbf{J}_1^{\phi} \right\rangle=
\frac{(d+2)}{(d+1)}I_1^{\phi}-I_0^{\dot{\phi}} \;.
\end{equation}
In order to form the simplest causal set expressions we can pick the strings $\vec{p}$ and $\vec{q}$, where only the lowest $k$ components are non-zero. Such strings are given in~\cite{Buck:2015oaa} so we will not repeat them here. If we sprinkle into a flat spacetime with a flat surface then $I_1^{\phi}=0$. In this case $\mathbf{J}_1^{\phi}$ is the causal set analogue of the normal derivative of $\phi(x)$ integrated across $\Sigma$.

Finally, we construct a causal set expression for the part in brackets in the third term on the RHS of~\eqref{eq:phi_expectation_final}. We define
\begin{equation}\label{general_causet_function_integral_phi_dotdot}
J_2^{\phi}\left[\mathcal{C},\mathcal{C}^+,\mathcal{C}^- ;\vec{p},\vec{q}\right]:=l^{d-3}d A^{\frac{3}{d}}\left( \sum_m p_m F_m^{\phi}\left[\mathcal{C}^- \right]
+  \sum_n q_n P_n^{\phi}\left[\mathcal{C}^+ \right] \right)\;,
\end{equation}
where $p_m$ and $q_n$ satisfy~\eqref{coefficient_relation}. The corresponding random variable, $\mathbf{J}_2^{\phi}$, can be shown to satisfy
\begin{equation}\label{mean_of_j2_to_get_integral}
\lim_{\rho\rightarrow\infty} \left\langle \mathbf{J}_2^{\phi}\right\rangle=\frac{1}{4(d+1)^2}I_2^{\phi}-
\frac{(2d+5)}{2(d+1)}I_1^{\dot{\phi}}+
\frac{1}{2}I_0^{\ddot{\phi}}\;.
\end{equation}

The same simple strings that were chosen towards the end of Section~\ref{a_causal_set_expression} can be chosen here to get simple causal set expressions that give the RHS of~\eqref{mean_of_j2_to_get_integral} as $\rho\rightarrow\infty$. Again, if we sprinkle into a flat spacetime with a flat surface, this expression will give something analogous to the second order normal derivative of $\phi(x)$ integrated across $\Sigma$.

\section{Conclusions and Further Work}

We have found a universal formula for the expansion of the volume of a small causal cone, up to $\mathcal{O}(T^{d+2})$. As the setup involves a hypersurface one might hope that the small volume formula can be used to derive the Hamiltonian formulation of General Relativity in the continuum, in a similar way to how Jacobson derives the Einstein equations using the volume of a small spacetime region in~\cite{Jacobson:2015hqa}.

The volume formula was used to find the limiting behaviour of the expectation values of certain random variables under a sprinkling process. With the limiting behaviour we were able to construct causal set expressions whose mean values corresponded, in the large density limit, to certain geometrical quantities relating to the hypersurface. We also looked at scalar fields on the causal set, and more expressions were constructed that related to normal derivatives of the scalar field.

The fluctuations of these causal set expressions were also studied in a simple case of a box in Minkowski spacetime. It was found that the fluctuations grew with the mean number of elements for dimensions less than $5$. If the causal set expressions are to be useful in $4$ dimensions then these fluctuations must be damped in some way. The use of an intermediate length scale in the Benincasa-Dowker-Glaser action seems like a promising approach that may dampen the fluctuations of the expressions found here. More work should be done to determine how the fluctuations scale in more complicated spacetimes, and in the case where a scalar field is included.

The geometric quantities that are encoded in the causal set expressions given here appear in the Hamiltonian formulation of General Relativity. Perhaps these causal set expressions can be used to formulate the dynamics of causal sets from a Hamiltonian perspective.

\section*{Acknowledgements}
I.J. is supported by the EPSRC. Many thanks to Michel Buck, Fay Dowker and Ollie Gould for their helpful suggestions towards this work.

\newpage
\bibliography{paper}
\bibliographystyle{jhep}

\end{document}